\documentclass[12pt,letterpaper]{article}
\usepackage[margin=1.2in]{geometry}
\usepackage{setspace}

\usepackage{epsfig, amsthm, amsmath, amssymb, latexsym, xpatch, xcolor, mathtools, verbatim, hyperref}
\usepackage[titletoc]{appendix}
\usepackage[mathscr]{euscript} 
\usepackage[normalem]{ulem} 
\usepackage{tabularx}
\usepackage{stmaryrd} 
\usepackage[margin=1cm]{caption}
\usepackage{float} 
\usepackage[margin=1cm, font={small,it}]{caption} 
\usepackage{wrapfig}
\usepackage{multicol} 

\definecolor{cardinal}{rgb}{0.8, 0.0, 0.0}

\usepackage{graphicx}

\newcommand{\op}{\begin{itemize}}
\newcommand{\ed}{\end{itemize}}

\newcommand{\opp}{\begin{quote}}
\newcommand{\edd}{\end{quote}}

\newcommand{\ope}{\begin{enumerate}}
\newcommand{\ede}{\end{enumerate}}

\newcommand{\im}{\item}

\newcommand{\PP}{\mathbb{P}}

\title{A Plea for History and Philosophy of Statistics and Machine Learning}

\author{Hanti Lin \\\\UC Davis \\ika@ucdavis.edu}

\date{} 

\begin{document}

\maketitle

\begin{abstract} 
\noindent Abstract: The integration of the history and philosophy of statistics was initiated at least by Hacking (1975) and advanced by Hacking (1990), Mayo (1996) and Zabell (2005), but it has not received sustained follow-up. Yet such integration is more urgent than ever, as the recent success of artificial intelligence has been driven largely by machine learning---a field historically developed alongside statistics. Today, the boundary between statistics and machine learning is increasingly blurred. What we now need is integration, twice over: of history and philosophy, and of two fields they engage---statistics and machine learning. I present a case study of a philosophical idea in machine learning (and in formal epistemology) whose root can be traced back to an often under-appreciated insight in Neyman and Pearson's 1936 work (a follow-up to their 1933 classic). This leads to the articulation of an epistemological principle---largely implicit in, but shared by, the practices of frequentist statistics and machine learning---which I call achievabilism: the thesis that the correct standard for assessing non-deductive inference methods should not be fixed, but should instead be sensitive to what is achievable in specific problem contexts. Another integration also emerges at the level of methodology, combining two ends of the philosophy of science spectrum: history and philosophy of science on the one hand, and formal epistemology on the other hand.
\end{abstract}

\newpage
\tableofcontents
\newpage

\section{Introduction}

Specialization in philosophy of science has given rise to diverse research programs, spanning a spectrum with two apparent poles: formal epistemology on the one hand, and history and philosophy of science on the other. The former is characterized by extensive use of mathematical tools, while the latter emphasizes historical case studies. Despite these differences, there is a growing need for synthesis. Let me explain.

While scientific inference is used by scientists in their empirical inquiries, it is also a subject of serious study by many, including not only epistemologists and philosophers of science, but also scientists themselves. Indeed, statistics---along with machine learning as its younger sibling---is a theoretical framework for studying scientific inference, developed {\em by} scientists and {\em for} scientists. Statistics and machine learning are scientific fields that study scientific inference.

As such, statistics and machine learning warrant philosophers' attention. Philosophical investigations into these fields, however, require two key methodologies: (i) the extraction and scrutiny of epistemological ideas from highly mathematical languages, especially the language of probability theory, and (ii) attention to the history and practice of the scientific fields in which those epistemological ideas were developed and applied. The former methodology corresponds to formal epistemology; the latter, to history and philosophy of science. Taking the lead, Hacking (1975, 1990), Mayo (1996), and Zabell (2005) already recognized the need to integrate these methodologies in their works on philosophy of statistics. I believe more philosophers should follow in their footsteps. The case for doing so is even stronger today, given recent advances in machine learning. It means taking the following questions seriously:
	\op
	\im {\em Descriptive Question}: Are there interesting epistemological views about scientific inference that remain largely implicit in the practices of statistics, machine learning, or other fields that study scientific inference (such as econometrics), and that still need to be articulated and made explicit?
	
	\im {\em Normative Question}: If so, how might those views be defended or challenged, thereby shedding new light on the epistemology of science---or even guiding scientific inquiry itself?
	\ed 
In this paper, I aim to offer partial answers to the above questions---if only to show how interesting and fruitful it can be to pursue an integrated history and philosophy of the integrated fields: statistics and machine learning.

To the descriptive question, I will answer in the affirmative by providing an example. Specifically, a series of historical case studies (Sections 3-5) will illustrate a recurring epistemological idea that has rarely been articulated but appears crucial to the foundations of statistics and machine learning. I call it {\em achievabilism}---the thesis that the correct standard for assessing non-deductive inference methods should not be fixed, but rather sensitive to what is achievable in specific problem contexts. 

This achievabilist idea can be traced at least as far back as Neyman and Pearson's (1936) seminal work, a follow-up to their much more frequently cited (1933) paper on hypothesis testing. I will present this episode of the history in Section 3. Since the achievabilist philosophy has mostly been practiced rather than articulated, it has often been taken lightly, even forgotten and reinvented---culminating in its recognition (though still not full articulation) in machine learning in the 1990s (Devroye, Gy\"{o}rfi, \& Lugosi 1996). This part of the history will be presented in Sections 4 and 5.

In response to the normative question, I will lay some groundwork for a future defense of achievabilism by exploring three philosophical applications:
	\op 
	\im One application addresses Norton's (2003, 2021) objection to formal theories of induction, advocating instead for theories that are formal but nonetheless {\em local}---sensitive to scientists' contexts of inquiry (Section 6).
	
	\im A second application recognizes Ockham's many {\em local} razor{\em s} in science and, despite their diversity, seeks to offer a unifying account (Section 7).
	
	\im A third (and final) application turns to traditional epistemology, challenging a common assumption about justified belief and inference---namely, that reliabilism implies externalism (Section 8).
	\ed 
I will then close with some of open questions in history and philosophy of statistics and machine learning (Section 9).

The challenge for me here---and for anyone else pursuing the history and philosophy of statistics and machine learning---is to conduct historical case studies by (i) delving into highly mathematical works, (ii) extracting the epistemological ideas therein, (iii) articulating those ideas in an accessible way for the broader philosophy of science audience, and (iv) evaluating both the ideas and the associated practices in light of the specific contexts in which they arose. In order to do all this for frequentist statistics, I feel it is necessary to begin with an entirely new, non-technical tutorial on frequentist statistics (next section)---if only to clarify the differing terminologies used by statisticians and philosophers, and to offer a roadmap of the concepts that will recur throughout the discussion (in fact, I will literally draw such a roadmap as Figure~\ref{fig-summary} in Section~\ref{sec-summary}).

\section{A Little Tutorial on Frequentist Statistics} 

Suppose we are interested in the unknown value of a certain quantity $\theta$---such as the bias of a coin, the proportion of people in a population who have a certain disease, or the chance of observing ``spin up'' in a quantum process. For concreteness, let's stick with the example of the coin bias. In frequentist statistics, alternative methods for making inferences about $\theta$ are evaluated using several key concepts, which appear as the titles of the subsections below and are explained therein.

\subsection{Physical Probabilities}\label{sec-probability}

At the possible world where the bias of the coin is $\theta = 0.3$, the physical probability of landing heads on each toss is equal to $0.3$:
	$$
	\begin{array}{lll}
	\PP_{0.3} \big( \textsf{Heads} \big) &=& 0.3 \,,\\
	\PP_{0.3} \big( \textsf{Tails} \big) &=& 1 - 0.3 \,.
	\end{array}
	$$
For simplicity, let the prescribed sample size be $n = 4$. Then the physical probability of obtaining a data sequence, say $(\textsf{Heads}, \textsf{Heads}, \textsf{Tails}, \textsf{Heads})$, can be decomposed under the assumption that the coin tosses are independent:
	\begin{eqnarray*}
	&& \PP_{0.3} 
	\big(\,
		(\textsf{Heads}, \textsf{Heads}, \textsf{Tails}, \textsf{Heads}) 
	\,\big)
\\
	&=& \PP_{0.3} \big( \textsf{Heads} \big) 
	\cdot \PP_{0.3} \big( \textsf{Heads} \big) 
	\cdot \PP_{0.3} \big( \textsf{Tails} \big)
	\cdot \PP_{0.3} \big( \textsf{Heads} \big)
	\end{eqnarray*}
Combining the above results, we have:
	\begin{eqnarray*}
	&& \PP_{0.3} \big((\textsf{Heads}, \textsf{Heads}, \textsf{Tails}, \textsf{Heads}) \big)
\\
	&=& 0.3 
	\cdot 0.3 
	\cdot (1 - 0.3)
	\cdot 0.3
	\end{eqnarray*}
This calculation procedure generalizes straightforwardly: for each (coarse-grained) possible world characterized by a specific coin bias $\theta$, the independence assumption determines a unique probability distribution $\PP_\theta$ over the $2^n$ possible data sequences, where $n$ is the number of coin tosses. Here, $\PP_\theta$ denotes the true distribution of physical probabilities in world $\theta$.

In the present context, the possible worlds correspond to the possible values of the parameter $\theta$---namely, the real numbers in the unit interval $[0, 1]$. The set of these possible values can be written as $\Theta = [0, 1]$. Statisticians call $\Theta$ the {\em parameter space} (for the given problem context), while philosophers may think of $\Theta$ as a space of coarse-grained possible worlds (a space that can be made more fine-grained by taking worlds to be pairs consisting of a parameter value and a data sequence).

At this point, you might be wondering what physical probabilities are, or how the term `physical probability' should be interpreted. This is an important metaphysical---or semantic---question. Perhaps physical probabilities are long-run relative frequencies (Neyman 1955, von Mises 1957), propensities (Popper 1959), Humean chances (Lewis 1981) or primitive physical states posited in science (Sober 2000: sec. 3.2). For focus, I will set aside this metaphysical/semantic debate and concentrate on epistemological issues. Suffice it to note that the probabilities used in {\em frequentist} statistics might---or might not---be best understood as long-run relative {\em frequencies}, since there are notable alternative interpretations. The term `frequentist' in `frequentist statistics' is therefore quite misleading---but unfortunately too entrenched to change.

\subsection{Competing Hypotheses, Inference Methods}

Continuing from the coin example, imagine that we are testing the following two hypotheses:
	\opp 
	$H_{0}$: $\theta \le 1/2$ (the bias of the coin is no more than 1/2).
	
	$H_{1}$: $\theta > 1/2$ (the bias of the coin is more than 1/2).
	\edd   
An inference method in this context is often called a {\em test}, a function that, given a data sequence, outputs a judgment: either rejecting the null hypothesis $H_0$, or not rejecting it (or perhaps even accepting it). 

More generally, an inference method is a function that maps any data sequence in a given family to a judgment about the hypotheses under consideration.

\subsection{Error}

Frequentist statisticians care about the avoidance of error. In a coarse-grained possible world (such as $\theta = 0.3$), when an inference method $M$ receives a data sequence as input (such as (\textsf{Heads}, \textsf{Heads}, \textsf{Tails}, \textsf{Heads})), $M$ returns a judgment as output---which may or may not be an error in that possible world. Here, an error is understood rigorously as an error {\em of} an inference method, {\em given} a data sequence, {\em in} a possible world.

In the context of frequentist hypothesis testing, there are two types of error. A {\em Type I error} occurs when the null hypothesis $H_0~(\theta \le 1/2)$ is rejected in a world where $H_0$ is true (such as the world $\theta = 0.3$). A {\em Type II error} occurs when $H_0$ is not rejected in a world where the alternative hypothesis $H_1$ is true (such as the world $\theta = 0.9$).

More generally, the concept of error can take different forms depending on the problem context. In problems of estimation (as opposed to hypothesis testing), where the competing hypotheses correspond to the possible values of a certain quantity, errors are typically quantifiable---often defined as the difference between the estimated value and the true value.

\subsection{Inductive Risk}

Once a conception of error is in place, a corresponding conception of inductive risk can be defined.

In the context of hypothesis testing, the inductive risk of an inference method $M$ at a possible world $\theta$ is typically defined as the physical probability that $M$ commits an error---either a Type I or a Type II error---under the true probability function $\PP_\theta$ in world $\theta$. Continuing from the coin example, in the world $\theta = 0.3$ (where $H_0$ is true), an inference method $M$ commits a (Type I) error exactly when we receive a data sequence $e$ as evidence such that $e$ prompts $M$ to reject the (true) hypothesis $H_0$---or in symbols, when $M(e) = $ ``Reject $H_0$''. The method's inductive risk at that world is then determined as follows:
	\begin{eqnarray*}
	&& \text{$M$'s inductive risk at world $\theta = 0.3$}
	\\
	&=& \text{the physical probability, at world $\theta = 0.3$, that $M$ commits a (Type I) error}
	\\
	&=& \PP_{0.3} \big( \text{$M$ commits a (Type I) error} \big) 
	\\
	&=& \PP_{0.3} \big( \text{we receive a data sequence that prompts $M$ to (erroneously) reject $H_0$} \big) 
	\\
	&=& \PP_{0.3} \big( \text{we receive a data sequence $e$ such that $M(e) = $ ``Reject $H_0$''} \big) 
	\\
	&=& \text{the sum of $\PP_{0.3}(e)$ over all data sequences $e$ such that $M(e) = $ ``Reject $H_0$''}
	\\
	&=& \sum _{e: \,M(e)\,=\, \text{``Reject\,$H_0$''}} \PP_{0.3} (e)
	\end{eqnarray*} 
Each term in the sum represents the physical probability of obtaining a particular data sequence in a given world---a quantity determined as explained in Section~\ref{sec-probability}.

There is another conception of inductive risk. When errors are quantifiable and take real numbers as values, the inductive risk of an inference method $M$ at a world $\theta$ can be defined as the expected value of the squared error at that world, where the expected value is a weighted average, with the weights given by the physical probabilities encoded in the probability function $\PP_\theta$. This conception of inductive risk is quite general: it subsumes the standard definition of inductive risk in hypothesis testing presented above, so long as the quantifiable error is defined as 1 when a Type I or Type II error occurs, and 0 otherwise.

Please note that statisticians do not typically use the term `inductive risk'. They simply use `risk' to refer to the concept in question, most often in contexts of estimation or regression---even though, as explained above, the concept readily generalizes to hypothesis testing. So the notion expressed by statisticians' `risk' is genuinely broad. Yet I choose to use a term more familiar to philosophers of science, `inductive risk', which nicely highlights that we are concerned with the risk incurred by inductive (i.e., non-deductive) inference.

\subsection{Risk-World Plot}

\begin{figure}
	\centering \includegraphics[width=0.95\textwidth]{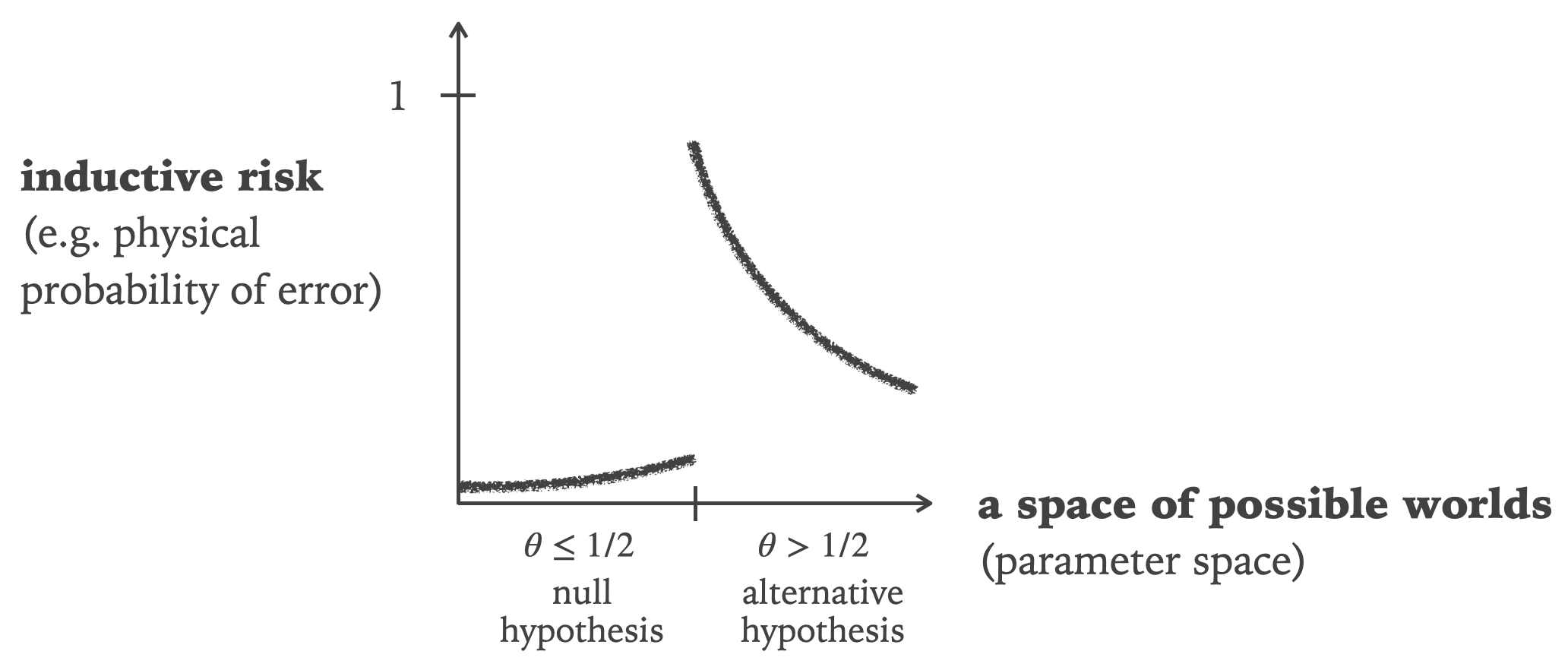}
	\caption{A risk-world plot of an inference method}
	\label{fig-plot}
	\end{figure}

Then, holding a particular sample size fixed, any inference method is associated with a risk-world plot, as illustrated in Figure~\ref{fig-plot}. The $Y$-axis represents inductive risk, while the $X$-axis represents the space of all possible worlds compatible with the background assumptions in the given context of inquiry.

Inductive risk can be plotted not only against possible worlds $\theta$ with a fixed sample size $n$, but also against pairs $(\theta, n)$, where both the possible world $\theta$ and the sample size $n$ vary. The resulting graph is a ``surface'' rising above a ``two-dimensional plane'' spanned by the axis of possible worlds and the axis of sample sizes.

\subsection{Frequentist Evaluative Standards}

Frequentist statisticians generally maintain that any inference method should be assessed by how its inductive risk varies across a range of possibilities---namely, the range of possible worlds compatible with the background assumptions in the given context. A frequentist standard for assessing inference methods (at a fixed sample size) is thus defined by examining the risk-world plots of those methods. Such a standard imposes a condition that excludes certain risk-world plots, and thereby rules out  the associated inference methods.

Here is an example of such an evaluative standard. Neyman and Pearson (1933) proposed a standard for hypothesis testing at a fixed sample size, known as {\em uniform maximum power} (UMP). The idea can be informally summarized as follows. An inference method $M$ is said to be uniformly most powerful at level $\alpha$ if and only if it satisfies the following two conditions:
	\op
	\im First, $M$'s inductive risk of committing a Type I error is guaranteed to be low---specifically, at most $\alpha$.
	\im Second, subject to the above constraint, $M$'s inductive risk of committing a Type II error is guaranteed to be minimized.
\ed
In this article, by `guarantee' I do not mean an unrestricted guarantee. Rather, it refers to a guarantee that holds under the background assumptions of a specific context of inquiry---that is, a guarantee defined by universal quantification over all possible worlds compatible with those assumptions.

\begin{figure}
	\centering \includegraphics[width=0.95\textwidth]{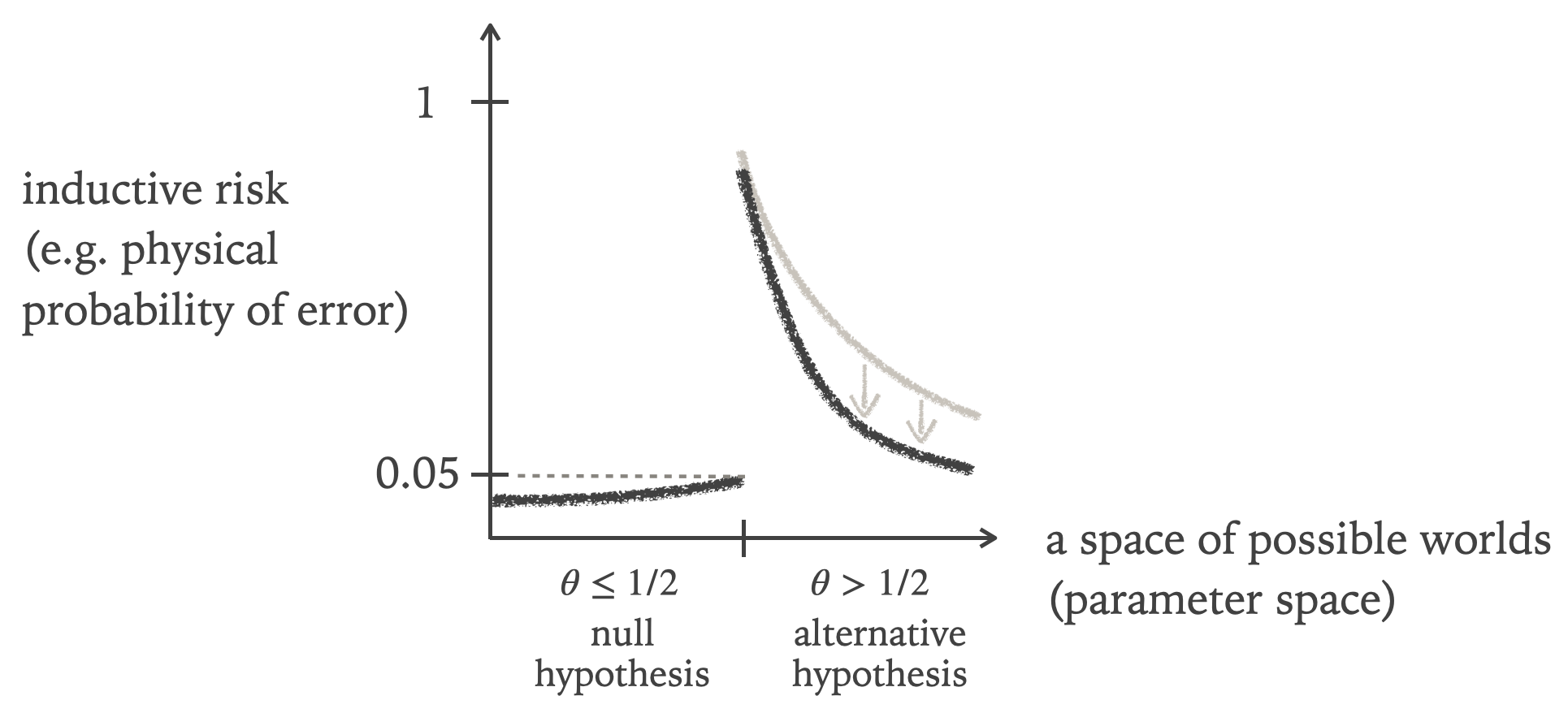}
	\caption{The idea of imposing the standard of UMP at level $5\%$}
	\label{fig-plot-np}
	\end{figure}

Now, making explicit the quantification over possible worlds, the first of the two guarantees can be more formally defined as follows (though still stated informally):
	\opp
	{\bf Definition (Low Level).} An inference method $M$ is said to have a {\em level} at (small) $\alpha$ iff, in every possible world (compatible with the background assumptions) where the null hypothesis $H_0$ is true, $M$ has an inductive risk of at most $\alpha$.
	\edd
By setting $\alpha$ to a specific small value, say 5\%, we obtain a candidate pool consisting of all tests with level 5\%. In terms of Figure~\ref{fig-plot-np}, this amounts to capping the left side of the risk-world plot at 5\%.
Neyman and Pearson (1933) then propose to narrow this pool further by requiring the second of the two guarantees, defined as follows:
	\opp 
	{\bf Definition (Uniformly Most Powerful).} An inference method $M$ is called {\em uniformly most powerful} (UMP) at level $\alpha$ iff the following two conditions are met:
	\op 
	\im[(1)] $M$ is an $\alpha$-level method.
	\im[(2)] In every possible world (compatible with the background assumptions) where the alternative hypothesis $H_1$ is true, the inductive risk of $M$ is less than or equal to that of any other $\alpha$-level inference method.
	\ed 
	\edd
In terms of Figure~\ref{fig-plot-np}, this amounts to minimizing the right side of the risk-world plot, subject to the $5\%$ cap on the left side. 

The above is an iconic example of a frequentist standard for assessing inference methods given a fixed sample size. More generally, a frequentist evaluative standard may also consider how an inference method's inductive risk varies not only across possible worlds but also across different sample sizes.

\subsection{Summary}\label{sec-summary}

\begin{figure}
	\centering \includegraphics[width=0.95\textwidth]{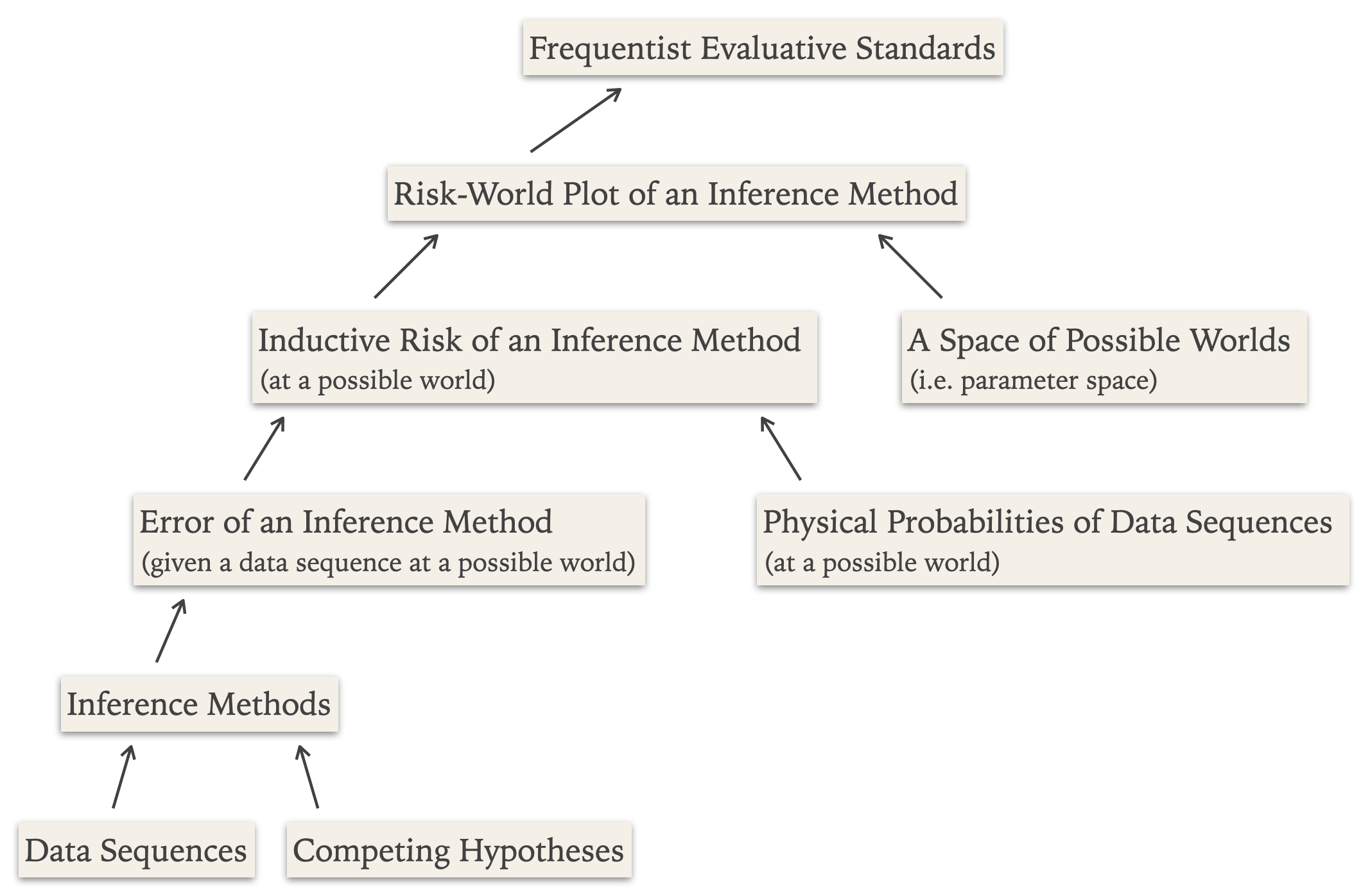}
	\caption{A roadmap of key concepts and their dependencies in frequentist statistics}
	\label{fig-summary}
	\end{figure}

Let me conclude this tutorial with a diagram that summarizes the dependencies among the concepts introduced above, as depicted in Figure~\ref{fig-summary}.\footnote{To clarify, this diagram intentionally omits some dependencies that are not important in the present context and that, if made explicit, would clutter the picture---such as the missing arrow from ``Data Sequences'' to ``Physical Probabilities of Data Sequences''.}
Perhaps this conceptual network has never been drawn in any textbooks or research articles. But if I am right, it expresses a common ground shared by many frequentist statisticians and philosophers sympathetic to frequentist statistics. It is also a common ground that highlights places where disagreements may arise, giving rise to several issues. One such source of controversy is interpretational, concerning how `physical probability' should be understood. As noted above, however, this is not our focus here.

Instead, this paper turns the spotlight to the top of this conceptual network: What frequentist standard should we use to evaluate inference methods? A general approach to this question was suggested by Neyman and Pearson three years after their classic 1933 paper. Enough background. The main show begins now.

\section{Genesis of Achievabilism}\label{sec-root}

In their 1933 classic paper, Neyman and Pearson studied this hypothesis testing problem: 
\begin{align*}
	\text{{\em Null Hypothesis} $H_{0}$:} & \quad\text{The bias of the coin $\le 0.5$.}
\\
	\text{{\em Alternative Hypothesis} $H_{1}$:} & \quad\text{The bias of the coin $> 0.5$.}
\end{align*}
This problem is {\em one-sided}, as the alternative hypothesis lays entirely on one side of the null hypothesis. For such one-sided problems, Neyman and Pearson (1933) succeeded in characterizing the inference methods that meet the standard of being uniformly most powerful at a low level (UMP). However, they soon discovered that the UMP standard is unachievable in the context of this problem:
\begin{align*}
	\text{{\em Null Hypothesis} $H_{0}$:} & \quad\text{The bias of the coin $= 0.5$.}
\\
	\text{{\em Alternative Hypothesis} $H_{1}$:} & \quad\text{The bias of the coin $\neq 0.5$.}
\end{align*}
This is a {\em two-sided} problem, as the alternative hypothesis $H_1$ includes values on both sides of the null hypothesis $H_0$. The question, then, is how to respond to such problem contexts, in which the high standard set by UMP is too high to be achievable. 

Three years after their 1933 classic paper, Neyman and Pearson proposed an answer---a gem that, I believe, deserves more attention in philosophy of statistics.

\subsection{Neyman \& Pearson, 1936}

One possible response to a two-sided problem is to insist on {\em universality}---that is, to adopt a single, lower standard applicable to all problem contexts, whether the problem at hand is one-sided, two-sided, or otherwise. However, this is {\em not} the approach recommended by Neyman and Pearson in their 1936 follow-up to their 1933 classic. 

What Neyman and Pearson (1936) proposed was a kind of context sensitivity. Here is the idea. In the context of a one-sided problem, the high standard of UMP is achievable, so anyone addressing that problem should aim for it. We should settle for a lower standard {\em only} when mathematical necessity demands it---that is, only when no inference method can meet the high standard of UMP in the given context. The two-sided problem is one such case. It is only in such contexts that Neyman and Pearson sought and accepted a lower standard. This contextualist stance was made abundantly clear in the very opening paragraph of their 1936 paper:
	\opp 
	In a number of recent publications [e.g., Neyman \& Pearson (1933)] we have discussed a method of approach to the problem of testing statistical hypotheses starting from a simple concept which may be expressed as follows: arrange your test so as to minimize the probability of errors. Since the errors involved in testing hypotheses are of two kinds, the problem requires further specification which may have different forms. One of these has led to the theory of uniformly most powerful tests, but as it has been found that in many situations [e.g. contexts involving a two-sided problem] a solution along these lines is impossible, it follows that {\em in such cases} the problem of minimizing the probability of errors must be specified in a {\em different} form. (Neyman \& Pearson 1936, p. 1, italics mine)
	\edd 
While the one-sided problem should still be addressed using the high standard of UMP, the two-sided case must be treated with a different, lower standard---one that is achievable in that context.


Neyman and Pearson's (1936) proposal for a lower standard is the condition of being {\em uniformly most powerful unbiased} (UMPU), which in effect inserts an intermediate step in the two-step definition of UMP: 
	\op
	\im {\em Step 1} (Old): Narrow down the candidate pool using the criterion of a low level---that is, by requiring that the inductive risk of committing a Type I error be guaranteed to be low (at most $\alpha$).
	\im {\em Step 2} (Newly Inserted): Further narrow down the candidate pool by requiring that the inductive risk of committing a Type II error be guaranteed not to be too high (e.g., at most $1 - \alpha$).
	\im {\em Step 3} (Old): Require that the inductive risk of committing a Type II error be minimized, subject to the above constraints---that is, minimized among the candidates remaining from the previous step.
	\ed
Note that the intermediate step acts as an additional filter, further narrowing the candidate pool before the final minimization step. If this intermediate step were omitted, we would return to the original standard of UMP, and the final step would minimize inductive risk over a larger set of inference methods, thereby imposing a more demanding standard. 

Here is a more formal presentation of Neyman and Pearson's lower standard. Start with this:
	\opp
	{\bf Definition (Unbiased).} An $\alpha$-level inference method is called {\em unbiased} iff its inductive risk of committing a Type II error is not too high---specifically, at most $(1 - \alpha)$---at every possible world (compatible with the background assumptions) where $H_1$ is true.
	\edd
The definition presented here is slightly stronger---but much easier to state in plain English---than Neyman and Pearson's original definition, which I include in the following footnote for reference.\footnote
	{
	An $\alpha$-level inference method $M$ is called {\em unbiased} according to Neyman and Pearson's original definition iff $M$'s lowest possible probability of accepting the alternative under the truth of the alternative is not too small---specifically, at least $\alpha^*$, where $\alpha^*$ is $M$'s lowest possible probability of accepting the null under the truth of the null (but if the minimum does not exist, use the greatest lower bound instead). To see why the definition I offer is stronger, note that $\alpha \ge \alpha^*$, given that $M$ is an $\alpha$-level inference method. While Neyman and Pearson's definition requires the relevant probability to be at least $\alpha^*$, my definition requires it to be at least $\alpha$---which is, in turn, at least $\alpha^*$. 
	} 
One more definition: 
	\opp 
	{\bf Definition (Uniformly Most Powerful).} Suppose that a candidate pool of inference methods, denoted $\mathscr C$, has been given. An inference method $M$ is called {\em uniformly most powerful} among the methods in $\mathscr C$ iff the following two conditions are met:
	\op 
	\im[(1)] $M$ is in class $\mathscr C$. 
	\im[(2)] In any possible world (compatible with the background assumptions) in which the alternative hypothesis $H_1$ is true, $M$'s inductive risk of committing a Type II error is less than or equal to that of any other method in class $\mathscr C$.
	\ed 
	\edd  
Then we obtain the following hierarchy of standards for hypothesis testing:
$$\begin{array}{l}
	\textit{UMP: Uniformly Most Powerful among $\alpha$-Level Tests}
\\
	\quad\quad\quad\; |
\\
	\textit{UMPU: Uniformly Most Powerful among Unbiased $\alpha$-Level Tests}
\\
	\quad\quad\quad\; |
\\
	\textit{A Low Level $\alpha$}
\end{array}$$
Let's step back for a moment before returning to the historical development.

\subsection{Distilling the Achievabilist Ideas}

Although Neyman and Pearson did not explicitly formulate a general epistemological thesis, it is not difficult to do so on their behalf:
	\opp
	{\em Weak Achievabilism}: The correct standard for assessing inference methods varies across problem contexts; it is not fixed universally, but depends on what standards are achievable in the context at hand.
	
	{\em Strong Achievabilism}: Moreover, an inference method is justified in a given problem context only if it meets the highest standard achievable in that context, provided that such a standard exists uniquely---pending a specification of the correct hierarchy of standards.
	\edd
Of course, Neyman and Pearson would insist on frequentist purity---that the correct hierarchy contains only frequentist standards, i.e., those that evaluate an inference method in terms of how its inductive risk varies across a range of possibilities. 

Now, what is a problem context? A problem context is specified by at least two factors:
\op
\im A set of competing hypotheses, representing the potential answers to the question posed and pursued within the context.
\im A space of possible worlds, or parameter values, representing the background assumptions taken for granted in the context and serving as the domain of quantification for the contextually relevant notion of guarantee.
\ed
When the sample size is held fixed, it constitutes an additional contextual factor. Otherwise, a possible world is understood as a possibility in which the sample size may increase indefinitely.

The general spirit is manifest in the active exploration of what can be achieved. If a given standard has been achieved, try raising the bar---possibly by exploring new standards. If the standards under consideration are shown to be unachievable, seek a lower, attainable one---again, possibly by exploring new standards. Or, in a slogan:
\begin{center}
\fbox{Look for what can be achieved; achieve the highest achievable.}
\end{center}

This is by no means the first time achievabilism receives a clear, general thesis statement. Such a statement has been offered at least by Lin (2025), who, however, focuses on the philosophical content and appears unaware of its historical root in the work of Neyman and Pearson, as well as its later development in machine learning (as discussed further below).

\section{Negligence in Statistics and in Philosophy}\label{sec-negligence}



Unfortunately, the achievabilist idea in Neyman and Pearson's work seems to have been largely neglected---even by Neyman and Pearson themselves when they write about philosophy. In his 1977 reflections on the philosophy of statistics, Neyman discusses only the higher standard of UMP, with no mention of the lower standard of UMPU, let alone the broader achievabilist perspective. Pearson's 1962 paper on statistical methodology similarly omits any reference to achievabilist ideas. To me, this is quite sad.

Worse, achievabilism continued to be overlooked by Lehmann, a student of Neyman's and author of the influential, now-classic textbook {\em Testing Statistical Hypotheses}. The first edition (1959) did motivate the lower standard of UMPU on page 125, presenting it as the result of lowering the bar from UMP, closely following Neyman and Pearson's (1936) rationale. A true achievabilist would then immediately ask: now that the bar has already been lowered once, might there be a problem context in which UMPU itself is unachievable, requiring a further lowering of the bar? Yet this natural question is not raised until much later---over a hundred pages on, at page 228. Note that I am not even talking about addressing the question, but merely about acknowledging it. The fourth edition (2022) shows no improvement on this front, retaining the same long delay. Here, the spirit of achievabilism seems to be taken lightly.


Worse still, the achievabilist idea in Neyman and Pearson's (1936) work has also largely eluded the attention of philosophers. In Hacking's {\em The Logic of Statistical Inference} (1965), only a single sentence hints at the idea of achievabilism, without further discussion: ``[W]hen no test is strictly UMP, one possible test is [UMPU] tests'' (p.~88). Even the most sustained philosophical defense and development of the Neyman-Pearson framework to date---Mayo's {\em Error and the Growth of Experimental Knowledge} (1996)---does not mention UMPU, let alone the broader achievabilist perspective.


While the origin of achievabilism in Neyman and Pearson's work on statistical hypothesis testing was largely neglected, the idea later re-emerged in other fields, especially in certain branches of theoretical computer science and formal epistemology. This may be because achievabilism is such a natural idea that it readily arose---independently, repeatedly, and without direct inspiration from Neyman and Pearson. This is the next arc of the story I want to tell.

\section{Repeated Reinventions of Achievabilism}\label{sec-repeated}

The idea of achievabilism was reinvented at least twice in theoretical computer science. The first reinvention occurred in the 1960s, within computability theory and algorithmic learning theory, and eventually made its way into a branch of philosophy known as formal epistemology. The second came in the 1980s, with the development of statistical learning theory, which has become one of the most influential approaches to the foundations of machine learning.

But before turning to the history of theoretical computer science, I would like to remain for a little while within the history of statistics---to highlight a few statisticians who took achievabilist steps but did not (yet) fully embrace the achievabilist spirit. In doing so, they unintentionally left some important work unfinished---work that would later be taken up by more serious achievabilists in theoretical computer science.

\subsection{Curve Fitting: Frequentist Statistics, 1980s-1990s}\label{sec-curve-fitting}

Imagine that we are working in a problem context of the following kind: we want to predict a house's price based solely on its location and floor area, or estimate a person's bone density based solely on their BMI (body mass index). More generally, we aim to draw a curve $y = f(x)$ on the $XY$-plane and use it for predictive purposes: once we observe that a unit (a house or a person) has a specific $X$-value, $x$, we use the curve $f$ to predict the corresponding $Y$-value of the unit, namely $f(x)$. Here is the key assumption: the primary goal in the present context is {\em prediction}. For simplicity, let $X$ and $Y$ be real-valued variables so that the $XY$-plane is literally a two-dimensional plane, which is easy to visualize (though in general, $X$ and $Y$ can be vector-valued).

Now the question is: how should we choose such a curve---a regression function---given a dataset represented as a scatterplot on the $XY$-plane? This is the central task of a {\em regression problem}.

\subsubsection{Model Selection: AIC vs. BIC}

A quite traditional approach to regression problems proceed in two steps:
\op 
\im {\em Step 1}: Pick a polynomial degree $d$ in light of the given dataset $D$ as a scatterplot on the $XY$-plane. This step serves to choose a model, i.e. a candidate pool: the class of polynomial functions of degree $d$.\footnote
	{
	Caveat: we don't have to use polynomials---we might instead consider classes of step functions or other families. One more caveat: the approach described above is just one way to do regression. There are alternative approaches that do not require choosing a model in the form of a function class---e.g., kernel regression, which still asks us to select something, namely a bandwidth.
	} 
\im {\em Step 2}: Pick, from that candidate pool, the function that has the best fit to the given dataset $D$. 
\ed 
This two-step procedure---model selection followed by choosing the best-fitting curve within the selected model---is a standard strategy in classical regression analysis. The second step is typically straightforward, as there is often a widely accepted measure of fit. The crux lies in the first step: model selection.

An obvious idea for model selection is Ockham's razor or the principle of parsimony: choose a model that strikes a balance between two desiderata, simplicity and fit to data. More specifically, consider two features of a model $M$: 
	\op 
	\im $M$'s simplicity, a virtue that can be understood in terms of the corresponding vice, complexity, typically measured by the number of parameters of $M$;
	\im $M$'s fit to data, defined as the best possible fit achieved by tuning the parameters of $M$.
	\ed  
Those two desiderata generally pull in opposite directions: fit to data can typically be improved by adding more parameters---that is, by using a more complex model. It is not immediately clear how this tension should be resolved, which has given rise to a now-classic competition between two model selection methods:
	\op 
	\im AIC, which may be called Ockham's blunter razor.
	\im BIC, which may be called Ockham's sharper razor, as it places greater weight on simplicity rather than fit. 
	\ed 
For a review and philosophical discussion of Ockham's razors in the context of curve fitting and model selection, see Forster and Sober (1994) and Sober (2008, Ch.~1).

We are now confronted with the task of assessing AIC and BIC as two competing inference methods. Which version of Ockham's razor should we adopt---and in what contexts? A somewhat achievabilist answer began to emerge from the statistician Shibata's work in the 1980s. But before introducing it, another brief tutorial is in order.

\subsubsection{A Tutorial on Regression via Point Estimation}

As a preliminary, let me offer a quick tutorial on regression, drawing a parallel with a more familiar inference task: point estimation. The task of point estimation can be described as follows:
	\op 
	\im In the problem context of {\bf point estimation}, we use data to produce a {\bf point} $a$ to estimate an unknown {\bf target quantity} $a^*$. 
	
	\im Whenever a specific {\bf point} $a$ is produced, the estimation error is measured by $|a - a^*|$, the difference between the {\bf numerical value} of $a$ and the {\bf numerical value} of the (unknown) target $a^*$. 
	
	\im In this context, we ask {\bf which number is the true value of the target quantity}, and would like to have an answer close to the true answer---close in the sense of a small estimation error. Any inference method in the present context, which returns a {\bf point} as an estimate whenever given a dataset, has an (unknown) inductive risk, defined as the (unknown) mean squared error---the expected value of the squared estimation error. An inference method is evaluated in terms of how its inductive risk varies across different possible worlds and sample sizes. 
	\ed 
In regression problems, we are basically doing the same thing, except that the target of estimation is not a single quantity but an entire curve or function (from $X$ to $Y$). To adapt the general description of point estimation to this case, we can copy the above text and modify just the bolded phrases accordingly, thereby obtaining a corresponding general description of regression problems.
	\op 
	\im In the problem context of {\bf regression}, we use data to produce a {\bf curve} or function $f: X \to Y$ to estimate an unknown {\bf target curve} $f^*$, {\bf being the curve having the best predictive power}.
	
	\im Whenever a specific {\bf curve} $f$ is produced, the estimation error is measured by the difference between the {\bf predictive power} of $f$ and the {\bf predictive power} of the (unknown) target $f^*$. 
	
	\im In this context, we ask {\bf which curve is the one having the highest predictive power}, and would like to have an answer close to the true answer---close in the sense of a small estimation error. Any inference method in the present context, which returns a {\bf curve} as an estimate whenever given a dataset, has an (unknown) inductive risk, defined as the (unknown) mean squared error---the expected value of the squared estimation error. An inference method is to be evaluated in terms of how its inductive risk varies across different possible worlds and sample sizes. 
	\ed 
The above presentation of regression problems leaves open how predictive power is defined; the technical details are unimportant and left to the following footnote.\footnote
	{
	The predictive power of a curve $f: X \to Y$ is often defined as the expected value of squared prediction error, i.e., the expected value of $|y - f(x)|^2$ with respect to the true probability density function $p(x, y)$ on the $XY$-plane.
	} 

What matters at this point is the parallel between point estimation and regression---a parallel that, in my view, deserves greater emphasis in the philosophy of statistics for the sake of conceptual clarity. One can begin with the more familiar case of point estimation and then understand regression as a natural generalization or variant thereof. This perspective explains why many commonly used evaluative standards in point estimation have been carried over---quite naturally---to the context of regression, such as these:
$$\begin{array}{l}
	\textit{Asymptotic Efficiency}
\\
	\quad\quad\quad\quad\quad |
\\
	\textit{Consistency (for Estimation)} 
\end{array}$$
Consistency for estimation is a guarantee (under the background assumptions of the given context) that the inductive risk of estimation converges to zero as the sample size increases indefinitely, whether the target of estimation is a single point or an entire function. Asymptotic efficiency raises the bar by imposing an additional requirement on the rate of convergence, though the details need not concern us here. When the target of estimation is not a point but a function, consistency for estimation is often referred to as {\em regression consistency}. 

Clarification on terminology: While the goal in the present context is to find a good predictive model, the term `consistency' is also often used relative to a different goal: that of identifying the true model. This alternative notion is known as {\em model selection consistency}, and it will not concern us in this section. The uses of the term `consistency' in statistics (and machine learning) can be quite confusing; context matters. End of this tutorial.

\subsubsection{The Rise of an Achievabilist Approach to AIC vs. BIC}

Now, which standard is achieved by AIC or BIC, and in what contexts? Thanks to Shibata (1981) and Shao (1997), we have the following interesting results:
	\op
	\im In problem contexts where the background assumption is that no model on the table is ``true'' (``true'' in the sense of containing the target---the best predictive curve), AIC, rather than BIC, achieves the higher standard of asymptotic efficiency (Shibata 1981).
	\im In problem contexts where the background assumption is that some model on the table is ``true'', BIC, rather than AIC, achieves the higher standard of asymptotic efficiency (Shao 1997).
	\ed
The underlying methodology aligns well with achievabilism. These results have motivated Shibata's and Shao's followers to investigate a more nuanced and realistic case---one in which the background assumption is weaker, remaining silent on whether any model under consideration is ``true'' (Yang 2005).

Asymptotic efficiency has thus become the standard criterion for assessing model selection methods in statistics, as employed by several widely cited survey articles (Arlot \& Celisse 2010; Cavanaugh \& Neath 2011; Ding, Tarokh, \& Yang 2018).


But perhaps because asymptotic efficiency has become too standardized in model selection within statistics, statisticians in this field seem to have largely failed to explore higher standards. Yet there is a clearly higher standard, well known in the theory of point estimation, called {\em uniform consistency}. This is consistency plus a short-run requirement: that there exists a particular sample size above which the inductive risk is guaranteed to be small---uniformly small across all possible worlds compatible with the background assumptions. 

So, an obvious extension of the above hierarchy is the following:
$$\begin{array}{l}
	\textit{Uniform Consistency (for Estimation)} 
\\
	\quad\quad\quad\quad\quad |
\\
	\textit{Asymptotic Efficiency}
\\
	\quad\quad\quad\quad\quad |
\\
	\textit{Consistency (for Estimation)} 
\end{array}$$
At this point, a true achievabilist would immediately ask: 
	\opp {\em Can we raise the bar in the context of a regression problem, reaching the high standard of uniform consistency?}
	\edd
Yet this question appears to have received little attention within the statistical community, as far as I can tell from the written record. This question was not formulated---let alone addressed---in the seminal works mentioned above: Shibata (1981) and Shao (1997). Nor was it posed in Claeskens \& Hjort (2008), a standard graduate-level textbook on model selection. In this respect, these statisticians seem not quite achievabilist enough. But perhaps I am wrong---perhaps the question was raised and discussed in conferences, even if not in published research papers or monographs. We need historians and philosophers of statistics to sort out what was actually the case.

Either way, here comes a dramatic part of the story: the achievabilist question posed above was actually answered shortly after Shibata's work in the 1980s, but not within statistics. The answer came from another field: theoretical computer science. The computer scientist Devroye (1982) answered the question in the negative, in a paper published in a computer science journal---one that remains rarely cited in the statistics literature, even to this day. It is time to turn to the story of Devroye, who took achievabilism further.

\subsection{Classification: Machine Learning, 1970s-2010s}\label{section-classification}

Classification is not very different from regression. Begin with our usual setup: the $X$-axis and $Y$-axis as two perpendicular real lines. Now, modify it slightly as follows. The $X$-axis, originally the space of real-valued inputs, undergoes an ``upgrade''---it becomes a space of possible images. On the other hand, the $Y$-axis experiences a ``downgrade''---the possible values of $Y$ are now restricted to be binary: $1$ means ``yes, the input is an image of a cat,'' and $0$ means ``no, it is not.'' The result is a {\em classification problem}---the problem of sorting out cat images from other images. A regression curve becomes a classifier in this context: it is still a function $f$ from $X$ to $Y$, but now $f$ outputs a value of the {\em binary} variable $Y$ whenever it is given a value of the {\em image-valued} variable $X$. 

Clarification: we might be interested in classifying things other than images---for instance, selecting an appropriate word to continue a given string of text. Consider the following string of words:
\opp
\texttt{Please write a 500-word essay on C. S. Peirce and his view on} \\\texttt{scientific inference. [End of Request] [Start of Response]} \\ \texttt{Charles Sanders Peirce is a philosopher known}
\edd
In this case, `\texttt{for}' seems to be a good candidate to continue the string. If this word continuation process is applied iteratively---turning into an iterated task of classification---you might eventually produce something that resembles an actual essay on Peirce's view of scientific inference. This is only the initial idea behind large language models (LLMs), but it conveys just how broad the scope of classification can be. That said, I'll stick with the cat image example to keep the ideas simpler and more concrete.

While a classifier is itself an algorithm, computer scientists are also interested in a different kind of algorithm: one that outputs a classifier when given a {\em training dataset}---a set of annotated images labeled with `1' or `0' by human annotators, much like a scatterplot on the $XY$ ``plane''. Such an algorithm is called a {\em learning algorithm}---an algorithm of a higher order, since its output, a classifier, is itself an algorithm. Learning algorithms for classification in machine learning are thus the counterparts of curve fitting methods for regression in statistics. As a result,  evaluative standards from one area can often be carried over to the other.

Interestingly, while the high standard of uniform consistency ought to be an obvious object of study, it has drawn attention primarily in the field of classification---and largely among theoretical computer scientists rather than statisticians. In a 1982 paper, the computer scientist Devroye proved that no learning algorithm achieves uniform consistency in problem contexts of classification where the background assumption is weak (imposing almost no constraint on the true, unknown probability distribution over the possible images of cats or non-cats). This impossibility result for classification can be easily carried over to regression.

Devroye's 1982 impossibility result was later reported in a monograph he coauthored with colleagues (Devroye, Gy\"{o}rfi, \& Lugosi 1996): {\em A Probabilistic Theory of Pattern Recognition}, a now-classic in the field known as {\em statistical learning theory}, which has been highly influential among machine learning theorists. Achievabilism is clearly practiced in this 1996 monograph, from which readers can readily reconstruct a hierarchy of standards for assessing learning algorithms:
$$\begin{array}{l}
	\mbox{{\em Uniform Consistency} (a short-run standard)} 
\\
	\quad\quad\quad\quad\quad |
\\
	\mbox{{\em Consistency with Certain Rates of Convergence}}
\\
	\quad\quad\quad\quad\quad |
\\
	\mbox{{\em Consistency} (a mere long-run standard)} 
\end{array}$$
This 1996 monograph also offers a systematic presentation of theorems of utmost importance to achievabilists, identifying which standards are achievable in which problem contexts---and which are not:
	\op 
	\im In a context where the set of competing classifiers to choose from is relatively small (in the technical sense of having a finite VC dimension), uniform consistency is achievable (Vapnik \& Chervonenkis 1974).
	\im  In a context where the set of competing classifiers to choose from is relatively large (with an infinite VC dimension), uniform consistency is unachievable (Devroye 1982) but consistency is achievable (Lugosi \& Zeger 1996).
	\ed 

Coincidentally, 1996 saw more than one monograph that painted a distinctly achievabilist picture. One came from machine learning: Devroye and colleagues' book, as we have seen. Another came from philosophy: Kelly's {\em The Logic of Reliable Inquiry} (1996), whose achievabilist heritage can be traced back to Putnam's (1965) work in theoretical computer science---but that is an arc of the story in philosophy I will tell later.


The bud of achievabilism continued to grow in machine learning. A clearly achievabilist picture finally received a textbook presentation for undergraduate students---in Shalev-Shwartz \& Ben-David's {\em Understanding Machine Learning: From Theory to Algorithms} (2014), which has become one of the most widely used textbooks in the field. Devroye's 1982 impossibility result, concerning a problem context of classification in which uniform consistency is unachievable, is prominently featured---memorably dubbed the {\em no-free-lunch theorem} for ease of recall and reference.

Readers of Shalev-Shwartz \& Ben-David's 2014 textbook can easily reconstruct a hierarchy of standards for assessing learning algorithms:
$$\begin{array}{ll}
	\mbox{{\em Uniform Consistency}} 
\\
	\quad\quad\quad |
\\
	\mbox{{\em Nonuniform Learnability}} \mbox{ (slightly stronger than mere consistency)}
\\
	\quad\quad\quad |
\\
	\mbox{{\em Consistency}}
\end{array}$$
(Terminological note: Shalev-Shwartz \& Ben-David uses the more descriptive term `convergence' instead of the less intuitive term `consistency'.) The exact definition of the intermediate standard---nonuniform learnability---is not important for now. What matters is this: in problem contexts where the set of competing classifiers is large, we already know from Devroye's (1982) no-free-lunch theorem that the high standard of uniform consistency is unachievable. Yet the low standard of consistency is achievable in those problem contexts, thanks to Lugosi \& Zeger (1996); in fact, they proved that it is achieved by a version of Ockham's razor for model selection in classification, known as {\em structural risk minimization} (SRM). 

An achievabilist would, naturally, ask whether Ockham's SRM razor can achieve a higher standard. Shalev-Shwartz \& Ben-David's 2014 textbook reported a positive answer: SRM achieves the intermediate standard of nonuniform learnability, thanks to Ben-David \& Jacovi (1993). This seems to me an important step for justifying SRM from an achievabilist point of view---a step that had long been overlooked in the philosophical literature on Ockham's SRM razor, such as Harman \& Kulkarni (2007) and Kelly \& Mayo-Wilson (2008). Philosophers have only recently begun to pay due attention to this result, thanks to Sterkenburg (2025, sec. 4.3.3).

It took thirty years for Devroye's 1982 bud of achievabilism to fully flourish in Shalev-Shwartz \& Ben-David's 2014 undergraduate textbook on machine learning. But when was the seed first sown? Did Devroye get the achievabilist idea from someone else? That, I don't know. In particular, I do not know whether Devroye was influenced by the achievabilist thought in Neyman \& Pearson (1936). Since Devroye was a theoretical computer scientists, it makes sense to try looking for possible roots within theoretical computer science.  

As it turns out, the earliest work I can find within theoretical computer science that presents a clearly achievabilist picture is due to Putnam (1965)---not in machine learning proper, but in a nearby subfield: computability theory. However, I can find no evidence that Devroye was influenced by Putnam's 1965 paper, which has been rarely cited in computer science.

Fortunately, Putnam was not only a computability theorist but also a philosopher, and the achievabilist spirit expressed in his 1965 paper began to lead a life of its own in philosophy. That is the next arc of the story I want to tell.

\subsection{Inductive Inference: Theoretical Computer Science and Formal Epistemology, 1960s-1990s}\label{section-inductive-inference}

Ignoring the (very important) requirement that any learning algorithm in use must be Turing computable, the concept of {\em decidability} in computability theory sets a high bar: it demands a guarantee that, as the amount of observational input increases indefinitely, the algorithm under assessment will eventually halt and, when it does, the output is the true hypothesis---the true answer to the question posed in context.

In his 1965 paper, Putnam was addressing a problem already known to be too hard to meet the high standard of decidability. His reaction was similar to Neyman and Pearson's back in 1936: the high standard should still be reserved for problems in which it is achievable; but for a (hard) problem in which the high standard is unachievable, let's lower the bar---an unmistakably achievabilist move---by tailoring the evaluative criterion to what is achievable in that particular problem context.

Let me reconstruct Putnam's move. Decompose decidability as the conjunction of two guarantees: 
	\op
	\im {\em Identification in the Limit} (a mode of convergence that is a non-stochastic counterpart of consistency in statistics and machine learning): a guarantee that, if the amount of observations were to increase indefinitely, the inference method would eventually output the true hypothesis and then never change its mind. 
	\im {\em Freedom from Error}: a guarantee that, if the amount of observations were to increase indefinitely, the inference method would never output a false hypothesis.
	\ed 
What Putnam did was, in effect, to drop the second guarantee and retain only the first as a lower standard. He also went further by defining an intermediate standard: identification in the limit (as a mode of convergence) together with a minimax requirement---namely, minimizing the worst-case number of mind changes incurred by the assessed algorithm across all possible worlds compatible with the background assumptions. The result was the following hierarchy of standards:
$$\begin{array}{ll}
	\mbox{{\em Decidability}} 
\\
	\quad\quad\quad |
\\
	\mbox{{\em Identification in the Limit} + {\em Minmax Mind Changes}} 
\\
	\quad\quad\quad |
\\
	\mbox{{\em Identification in the Limit}} 
\end{array}$$
Although Putnam painted a clearly achievabilist picture, all of this was unfortunately framed within computability theory, and his hierarchy of standards was applied to assess algorithms for computing arithmetic functions, rather than for explicitly learning about the empirical world. As a result, readers could easily fail to see the connection to non-deductive inference in the empirical sciences.

Good news: almost the same idea was developed by the computer scientist Gold (1967), who applied it to evaluate algorithms aimed at explicitly learning something about the empirical world---specifically, learning the grammar of a language by observing utterances from native speakers. Although Gold studied only two standards in Putnam's hierarchy---the highest and the lowest---his work inspired a generation of theoretical computer scientists who embraced the achievabilist spirit and went on to explore intermediate standards as well. For recent surveys, see Jain \& Stephan (2017) and Case \& Jain (2017).

But there is also bad news: although the achievabilist line of research stemming from Gold gave rise to an established subfield in theoretical computer science---known as {\em algorithmic learning theory} or {\em inductive inference}---it is somewhat inactive in today's computer science.

Fortunately, there is more good news: Putnam was not only a computability theorist but also a philosopher. Because of this, the achievabilist legacy of Putnam and Gold from the 1960s was inherited and sustained by a small group of formal epistemologists in philosophy, who established the field known as {\em formal learning theory}. Here is a classic problem context studied in this field, which we may call the {\em raven problem}: we pose the question whether all ravens are black, with ``Yes'' and ``No'' as the two competing hypotheses, under the background assumption that counterexamples to the general hypothesis (``Yes''), if they exist, would eventually be observed if the number of observations were to increase indefinitely. Building on Kelly's (1996) monograph {\em The Logic of Reliable Inquiry}, Schulte (1999) and Kelly (2004) proved some important results to the following effect: in the raven problem and many other problems of testing deterministic hypotheses, Putnam's intermediate standard (or a slight variant) is the {\em highest achievable}, and it is achieved {\em only by} inference methods that follow a specific type of Ockham's razor. This type may be called Ockham's {\em Popperian} razor, as it employs a conception of simplicity that Popper (1959$b$) explored.\footnote
	{	
	More specifically, Popper (1959$b$) proposed a falsifiability-based conception of simplicity: a hypothesis is said to be simpler than another in the Popperian sense iff the former is more falsifiable---that is, whenever the former is falsified by empirical evidence, so is the latter, but not the other way around. For example, it is not hard to check that ``all ravens are black'' is simpler than its negation in the Popperian sense. 
	}
The idea of achievabilism is clearly implemented and practiced in Kelly's and Schulte's justifications of Ockham's Popperian razor in the intended local contexts.

Even though Kelly and Schulte still stopped short of offering a clear thesis statement of achievabilism, the idea was made more salient than ever. For the first time in the history of achievabilism, readers were not left to reconstruct a hierarchy of standards from the text: Kelly (1996) explicitly included a figure presenting such a hierarchy (p.~65), making the achievabilist structure of the theory visually and conceptually accessible.

Yet perhaps due to the absence of a clear thesis statement, the idea of achievabilism was not always taken seriously enough in Kelly's 1996 monograph. Indeed, he offered a somewhat Kantian view on the standard of identification in the limit: intentionally or not, he wrote as if this standard were a necessary condition for the very possibility of scientific knowledge (Kelly 1996, pp.~36-37). This amounts to claiming that identification in the limit---a {\em non-stochastic} mode of convergence---is the minimum qualification for justified scientific inference across all problem contexts. Such a move, however, should be resisted by achievabilists, as it excludes statistics and machine learning, which focus on standards such as consistency and other {\em stochastic} modes of convergence.


It was only recently that philosophers in the tradition of formal learning theory began to formulate a clear, general thesis statement of achievabilism, in part to avoid unnecessary or unintended commitments---as in Lin (2025), who, unfortunately, appeared unaware of achievabilism's root in statistics and its development in machine learning.

This concludes all the historical case studies I set out to examine. It is now time to shift from history to philosophy---to explore the philosophical applications of achievabilism.


\section{Philosophical Applications of Achievabilism}\label{sec-app}

I believe that achievabilism is interesting because of its potential for philosophical applications, three of which will be presented below.

\subsection{Against Norton Against Formal Theories of Inference}

Achievabilism, as presented above, grew out of an epistemological tradition that sought to develop theories of scientific inference with the aid of formal tools---such as symbolic logic and probability theory. Within this broadly formality-focused tradition, we also find familiar views such as the subjective and objective varieties of Bayesianism, which typically uphold a normative standard only if it is universally applicable---for example, the norm that one's degrees of belief should always satisfy the formal axioms of probability.

Such theories---formal theories of scientific inference---have been the targets of criticism, notably by Norton (2003, 2021). Norton argues that a formal theory of scientific inference must be universally applicable in the following sense: it must require that all scientific inferences, in all contexts, conform to a fixed form that serves as a universal standard---one size fits all. His thesis can be captured by the following slogan:
	\opp 
	{\em Norton's Implication}. Formality implies universality.
	\edd  
Given this implication, Norton applies {\em modus tollens} to argue against formality. That is, he argues that induction should be local and context-sensitive, so there should not be a universal standard pretending to be ``one size fits all'' in a theory of scientific inference. Then, by Norton's implication, he concludes that a theory of scientific inference should not be formal.

However, Norton's implication may be challenged. Its plausibility hinges crucially on how formality is understood. If formality means only the use of mathematical tools---such as symbolic logic or probability theory---to formalize evaluative standards and to present them in schematic forms, then Norton's implication does not hold. In fact, achievabilism provides a counterexample: it is a formal theory in this sense, but it explicitly rejects the demand for a universal standard. Let me explain.

Think of, for example, the standard of uniform consistency, which was discussed above for regression and classification. Now is the moment when we (finally!) need to examine its formal definition (though still stated as informally as possible):
\opp 
{\bf Definition (Uniform Consistency).} An inference method $M$ is said to achieve {\em uniform consistency} iff, for any threshold of inductive risk $\epsilon > 0$, there exists a natural number $N$ such that $M$ has an inductive risk less than $\epsilon$ given any sample size greater than or equal to $N$ in every possible world on the table, where
\op 
\im a possible world on the table is one compatible with the {\bf background assumption}, 
\im and the inductive risk is the expected value of the error incurred when inferring a falsehood among the {\bf competing hypotheses}.
\ed 
\edd
This evaluative standard is {\em formal} in two senses. First, it is {\em formalizable mathematically}. Second, it is {\em schematic}---the bolded elements in the definition can be freely replaced. Yet, it is precisely because of this formal, schematic character that these bolded elements can be replaced by contextual factors, being sensitive to one's specific context of inquiry. Moreover, the criterion of uniform consistency need not be upheld as a universally applicable standard. In fact, most achievabilists hold that, in problem contexts where uniform consistency is unachievable, that standard simply doesn't apply, and the bar must be lowered to identify the standard that applies---the highest achievable one.

In short: formality, when properly construed, does {\em not} imply universality. A theory can be formal and yet context-sensitive, and achievabilism is a clear example of this---an example that grew out of the actual practice of scientists who studied scientific inference.

To be sure, Norton's implication might be saved by restricting its scope---by strengthening the antecedent to invoke a narrower sense of formality. I suspect that, in that case, achievabilism would no longer be categorized as a ``formal'' theory of scientific inference, but it should still play an important role: to help us sharpen Norton's implication and identify a plausible version of it.

From a hindsight that is familiar to many philosophers, the possibility of using formal tools to defy universality and embrace context-sensitivity should come as no surprise. Since at least Kaplan's works (1989$a$, 1989$b$) in formal semantics, linguists and philosophers of language have been using formal tools to study context-sensitivity in natural languages, leading to an explicit, formal modeling of how an expression's denotation (or semantic value) varies across different contexts. There is no reason why theorists of induction cannot do the same. In fact, Neyman and Pearson already started it back in 1936 when they pioneered the idea of achievabilism.


\subsection{Unified Justification of Ockham's Many Local Razors}

If you recall, tentative justifications of {\em local} versions of Ockham's razor have emerged from three of the four historical case studies above: 
	\ope  
	\im Ockham's (blunt) AIC razor vs.\ (sharp) BIC razor in contexts of regression and model selection (Section \ref{sec-curve-fitting});
	\im Ockham's SRM razor in contexts of classification (Section \ref{section-classification}); 
	\im Ockham's Popperian razor in contexts of testing deterministic hypotheses (Section \ref{section-inductive-inference}). 
	\ede
A shared pattern emerges. In each case, there is a hierarchy of standards (possibly a fragment of a larger one) that takes the following form:
$$\begin{array}{ll}
	\quad\quad\quad\quad\quad \vdots
\\
	\mbox{{\em A Relatively High Standard}}
\\
	\quad\quad\quad\quad\quad |
\\
	\mbox{{\em An Intermediate Standard}} 
\\
	\quad\quad\quad\quad\quad |
\\
	\mbox{{\em A Mere Long-Run Standard about Inductive Risk}} 
\\
	\quad\quad\quad\quad\quad \vdots
\end{array}$$
And, in each case, we always find a class of problem contexts in which the relatively high standard is too strong to be achievable, and the mere long-run standard is too weak to imply anything of real interest. Yet an intermediate standard is explored, precisely defined, and shown to be connected to Ockham's razor in interesting ways. Let me give a quick summary below.

First of all, the result is particularly strong in the case of testing deterministic hypotheses (as in formal learning theory): the intermediate standard therein is achievable, and achieved only by Ockham's Popperian razor. In the case of regression and model selection in statistics, the result is not as strong, but it still helps resolve a significant debate: the intermediate standard therein is achievable and serves to rule out at least one of Ockham's two notable competing razors---AIC vs. BIC---depending on the background assumptions in context. Then, by appealing to the normative principle ``Achieve the highest achievable,'' we arrive at context-sensitive justifications of Ockham's local razors in different contexts---unified under achievabilism.

Turning to the case of classification in machine learning, the available result is not yet as strong as in the previous two cases: the intermediate standard is achievable and rules in the SRM version of Ockham's razor, but further work is needed to determine whether it rules out any interesting alternative methods. 

There are two things to like about the achievabilist style of justification. The first is {\em unification}. It strikes me as the most unifying approach we currently have for justifying Ockham's many local razors. It does not make the na\"{i}ve assumption that the world is simple, nor does it postulate the primitive normativity of Ockham's razor (as in Swinburne 1997). Instead, it seeks to justify Ockham's local razors within their respective local contexts, guided by a common principle: strive for the highest standard achievable in context---a standard concerned with minimizing, controlling, or diminishing inductive risk.

The second thing to like about the achievabilist style of justification is its recent momentum. In statistical hypothesis testing, Genin (2018) uses an achievabilist strategy to justify a type of Ockham's razor that explains why we ought to prioritize controlling errors for certain hypotheses---namely, the simpler ones, often called null hypotheses. This addresses a longstanding complaint about the Neyman-Pearson theory. Likewise, in a branch of machine learning known as causal discovery (or causal structure learning), Lin (2019) and Lin \& Zhang (2020) explicitly adopt achievabilism to justify what might be called Ockham's causal razor: ``Don't infer causation between two variables unless you believe there is association between them.'' Or more succinctly: ``No causation without association.''

All this is quite sketchy---much work is still needed for a detailed account, defended against alternative approaches to Ockham's razor. But I hope this is enough to suggest how it is possible to develop a unified justification of Ockham's many local razors.


\subsection{The Possibility of Internalist Reliabilism}

While the preceding two applications pertain to philosophy of science, the following one brings us to traditional epistemology.

There has long been an impression in traditional epistemology that reliabilism implies externalism. Relatedly, Otsuka (2023, sec.~3.3) argued that frequentist statistics---due to its reliabilist nature---is inherently externalist. I disagree. I believe that frequentist statistics can plausibly be understood as a form of internalist reliabilism---internalist in the conventional sense, yet reliabilist in some unconventional ways, which become salient when viewed through the achievabilist lens. Let me explain.

According to the conventional version of reliabilism (Goldman 1979), the justification of an inference method $M$ depends solely on the actual reliability that $M$ possesses. This actual reliability---or unreliability---may be measured by the actual inductive risk of $M$: the inductive risk that $M$ has at the actual world. This factor of justification---actual inductive risk---is external in the sense that it is not always possible, even in principle, for an agent to come to recognize the actual inductive risk of their method through mere reflection or introspection. And it is in this sense that conventional reliabilism is undeniably externalist.

In contrast, the implementations of achievabilism across statistics, machine learning, and formal epistemology are internalist. The crucial reason is that the evaluative standards under consideration---such as uniform consistency---do not really rely on the inductive risks at the actual world. Instead, they refer to the inductive risk of the assessed method at each of the possible worlds compatible with the background assumptions that one makes in one's context of inquiry---whether or not the actual world turns out to be one of those worlds. Such inductive risks are something that one can, in principle, examine by mere reflection or introspection alone. Let me highlight the elements that make this approach internalist:
	\opp  
	When an evaluative standard refers to inductive risk, it is always the {\em \uline{inductive risk}}$_{\,(c)}$ of an inference method at a possible world {\em \uline{compatible}}$_{\,(b)}$ with one's {\em \uline{background assumptions}}$_{\,(a)}$, where the three underlined elements $(a)$-$(c)$ are all internalist:
	\op 
	\im[$(a)$] One's {\em background assumptions} are in principle something that one can recognize in one's context of inquiry by reflection alone.
	
	\im[$(b)$] It is a mathematical (or even logical) question as to whether a possible world is {\em compatible} with one's background assumptions; qua mathematical question, its answer can in principle be obtained and recognized by reflection alone (barring the issues arising from the incompleteness of mathematics).
	
	\im[$(c)$] It is a mathematical question as to what the {\em inductive risk} of an inference method is at a given sample size and a given possible world (be it the actual world or not); qua mathematical question, its answer can in principle be obtained and recognized by reflection alone  (barring, again, the issues arising from the incompleteness of mathematics). 
	\ed 
	\edd 

While the implementations of achievabilism across statistics, machine learning, and formal epistemology are clearly internalist, as seen above, they are reliabilist nonetheless---in two important senses. First, they are reliabilist in that they assign a central role to inductive risk when defining evaluative standards. Second, they are reliabilist in taking the pursuit of reliability seriously: they strive to enforce the highest achievable standard for minimizing, controlling, or diminishing inductive risk across a range of (internally accessible) possible worlds.

At this point, some might insist that the label `reliabilism' be reserved only for what I have called conventional reliabilism. I do not intend to engage in terminological disputes. Yet there is an important philosophical message here: the achievabilist framework, as developed in statistics, machine learning, and formal epistemology, can be understood as (i) being internalist and, simultaneously, (ii) being reliabilist in the two senses just discussed---especially in the second, achievabilist sense of reliabilism, which urges an (internalist) serious pursuit of reliability that, I believe, is arguably one of the pillars in the foundations of frequentist statistics and machine learning. The possibility of combining (i) and (ii) is easily overlooked if we insist too strongly on using a language that saves the truth of `reliabilism implies externalism'.

So, my aim is not to stretch the meaning of `reliabilism' merely to block the implication from reliabilism to externalism. Rather, the goal is to gain a deeper understanding of frequentist statistics and machine learning---fields that represent some of the most sustained and rigorous attempts by scientists to develop an epistemology for their own use.

\section{Closing}

I have presented a story of the historical development of achievabilism: its root in statistics (Section \ref{sec-root}), and its repeated reinvention in statistics, machine learning, and other areas of theoretical computer science (Sections \ref{sec-negligence} and \ref{sec-repeated}).

I have also sketched three applications of achievabilism (Section \ref{sec-app}). A full-fledged defense of achievabilism as an epistemological framework has to be reserved for future work.\footnote
	{That said, Lin (2025) has outlined a plan to defend achievabilism, or achievabilist convergentism, and compare it with three more well-known approaches in the epistemology of scientific inference: Bayesianism, instrumentalism, and explanationism (based on inference to the best explanation).}
Yet, like any philosophical theory, the case for achievabilism must rest in part on the strength of its applications. By focusing on some notable applications here, I hope to take a first step toward such a defense.

The lesson I want to draw is that there are interesting epistemological views that are implicitly practiced in statistics and machine learning but remain in need of explicit articulation for critical assessement. Achievabilism is one such example, and I believe there must be others. Perhaps some of these can be uncovered by addressing pressing open questions such as the following:

{\bf Open Question I (from Statistics).} The greatest divide in philosophy of statistics is between frequentist and Bayesian approaches (although the present paper has largely ignored the development of Bayesianism). However, hybrid methods have been developed since the 1960s (Freedman 1963). Is the divide real? Is the hybrid a coherent middle way, or just a mixed bag of incoherent ideas?

{\bf Open Question II (from Machine Learning)}. In the section on classification, I discussed the development of statistical learning theory as an instance of achievabilist thinking---an approach that has long aspired to serve as the (frequentist) epistemological foundation of machine learning. However, it has recently been argued---most notably by computer scientists such as Moritz Hardt---that statistical learning theory has little to do with the actual advancement of one of the most fascinating and successful branches of modern machine learning: deep learning. In his keynote speech at the 2024 ICLR (International Conference on Learning Representations), Hardt proposed that deep learning requires a new epistemological foundation, which he dubbed the {\em emerging science of benchmarks} (\url{https://iclr.cc/virtual/2024/invited-talk/21799}). Is statistical learning theory mistaken? Does the emerging science of benchmarks complement, or compete with, statistical learning theory?

{\bf Open Question III (from Econometrics)}. While statisticians had been largely dismissive about causal inference throughout most of the history of statistics, the 2021 Nobel Prize in Economics recognized a theory of causal inference---the potential outcomes framework---which had its early development (Rubin 1974) much better received in econometrics rather than in statistics. Was there a philosophical view, held either explicitly or implicitly by some statisticians at the time, that resisted the very possibility of causal inference? Did practitioners of causal inference develop any philosophical views in response? Or what would constitute a good response?


All this suggests that scholars in history and philosophy of science (HPS) and those in formal epistemology should join forces to uncover epistemological positions that are practiced---but not yet explicitly articulated---in the scientific fields that study scientific inference. This includes statistics and machine learning and their sister fields, such as econometrics (a subfield of economics that is continuous with statistics) and phystatistics (the subfield of high-energy physics that is devoted to statistical data analysis).\footnote
	{The term `phystatistics' seems to be a relatively recent coinage; see B. Nachman's opinion article ``The Rise of the Data Physicist'' published here: \url{https://www.aps.org/apsnews/2023/10/rise-of-data-physicist}} 
The collaboration between HPS and formal epistemology is vital for enriching our space of possible epistemological positions and for putting us in a better position to select and defend a view that is both philosophically robust and scientifically grounded. This collaborative task is all the more urgent given the recent advances in machine learning.

I hereby make a plea for one possible conception of \textbf{HPS$^+$}: it abbreviates \textbf{H}istory and \textbf{P}hilosophy of \textbf{S}tatistics \textbf{plus} Machine Learning, and it also stands for \textbf{H}istory and \textbf{P}hilosophy of \textbf{S}cience \textbf{plus} Formal Epistemology. Hacking (1975, 1990), Mayo (1996), and Zabell (2005) have taken the lead. To follow in their cross-disciplinary footsteps, we need greater integration of methodologies between HPS and formal epistemology---combining attention to the practice of science in context with the pursuit of rigorous epistemology grounded in probability theory and other mathematical tools. We also need better tutorials and pedagogical strategies to lower the barrier for students moving between these two areas.

\subsection*{Acknowledgements} I thank Reuben Stern, DJ Arends, Gordon Belot, Jim Joyce, Jun Otsuka, and Conor Mayo-Wilson for valuable discussions. I am particularly grateful to Michela Massimi and Kevin K. T. Kelly, who sowed the seeds of HPS and formal epistemology in me, respectively---without which this paper would not have been possible.


\subsection*{References}

\begin{description}
\im Arlot, S., \& Celisse, A. (2010). A survey of cross-validation procedures for model selection. {\em Statistics Surveys, 4}, 40-79.

\im Ben-David, S., \& Jacovi, M. (1993). On learning in the limit and non-uniform $(\varepsilon, \delta)$-learning. In {\em Proceedings of the Sixth Annual Workshop on Computational Learning Theory} (pp. 209-217).

\im Case, J., \& Jain, S. (2017). Connections Between Inductive Inference and Machine Learning. In C. Sammut \& G. I. Webb (Eds.), {\em Encyclopedia of Machine Learning and Data Mining} (pp. 261-272). Springer, Boston, MA.

\im Cavanaugh, J. E., \& Neath, A. A. (2019). The Akaike information criterion: Background, derivation, properties, and refinements. {\em Wiley Interdisciplinary Reviews: Computational Statistics, 11}(3), e1460.

\im Claeskens, G., \& Hjort, N. L. (2008). {\em Model Selection and Model Averaging}. University Cambridge Press.

\im Devroye, L. (1982). Any discrimination rule can have an arbitrarily bad probability of error for finite sample size. {\em IEEE Transactions on Pattern Analysis and Machine Intelligence, PAMI-4}(2), 154-157.

\im Devroye, L., Gy\"{o}rfi, L., \& Lugosi, G. (1996). {\em A Probabilistic Theory of Pattern Recognition}. Springer.

\im Ding, J., Tarokh, V., \& Yang, Y. (2018). Model selection techniques: An overview. {\em IEEE Signal Processing Magazine, 35}(6), 16-34.

\im Forster, M., \& Sober, E. (1994). How to tell when simpler, more unified, or less ad hoc theories will provide more accurate predictions. {\em The British Journal for the Philosophy of Science, 45}(1), 1-35.

\im Genin, K. (2018). The topology of statistical inquiry. PhD dissertation, {\em Carnegie Mellon University}.

\im Gold, E. M. (1967). Language identification in the limit. {\em Information and Control, 10}(5), 447-474.

\im Goldman, A. I. (1979). What is justified belief?. In G. S. Pappas (Ed.), \textit{Justification and knowledge: New studies in epistemology} (pp. 1-23). Springer Netherlands.

\im Hacking, I. (1965). {\em Logic of Statistical Inference}. Cambridge University Press.

\im Hacking, I. (1975). {\em The Emergence of Probability: A Philosophical Study of Early Ideas about Probability, Induction and Statistical Inference.} Cambridge University Press.

\im Hacking, I. (1990). {\em The Taming of Chance.} Cambridge University Press.

\im Harman, G., \& Kulkarni, S. R. (2007). {\em Reliable Reasoning: Induction and Statistical Learning Theory}. MIT Press.

\im Jain, S., \& Stephan, F. (2017). Inductive Inference. In C. Sammut \& G. I. Webb (Eds.), {\em Encyclopedia of Machine Learning and Data Mining} (pp. 642-648). Springer, Boston, MA.

\im Kaplan, D. (1989$a$). Demonstratives. In J. Almog, J. Perry, and H. Wettstein (eds.), {\em Themes from Kaplan} (pp. 481-563). Oxford: Oxford University Press.

\im Kaplan, D. (1989$b$). Afterthoughts. In J. Almog, J. Perry, and H. Wettstein (eds.), {\em Themes from Kaplan} (pp. 565-614). Oxford: Oxford University Press.

\im Kelly, K. T. (1996). {\em The Logic of Reliable Inquiry}. Oxford University Press.

\im Kelly, K. T. (2004). Justification as truth-finding efficiency: How Ockham's razor works. {\em Minds and Machines, 14}, 485-505.

\im Kelly, K. T., \& Mayo-Wilson, C. (2008). Review of Harman \& Kulkarni (2007) Reliable Reasoning: Induction and Statistical Learning Theory. {\em Notre Dame Philosophical Reviews}.

\im Lehmann, E. L. (1959). {\em Testing Statistical Hypotheses} (1st ed.). John Wiley \& Sons.

\im Lehmann, E. L., Romano, J. P., \& Casella, G. (2022). {\em Testing Statistical Hypotheses} (4th ed.). Springer.

\im Lewis, D. (1981). A subjectivist's guide to objective chance. In Harper, W. L., Stalnaker, S, \& Pearce, G. (eds.) {\em Ifs: Conditionals, belief, decision, chance and time} (pp. 267-297). Springer Dordrecht.

\im Lin, H. (2019). The hard problem of theory choice: A case study on causal inference and its faithfulness assumption. {\em Philosophy of Science, 86}(5), 967-980.


\im Lin, H. (2025). Convergence to the truth. In K. Sylvan, J. Dancy, E. Sosa, \& M. Steup (Eds.), \textit{The Blackwell companion to epistemology} (3rd ed.). Wiley Blackwell.

\item Lin, H., \& Zhang, J. (2020). On learning causal structures from non-experimental data without any faithfulness assumption. {\em Proceedings of Machine Learning Research, 21}(1), 1-36.


\im Lugosi, G., \& Zeger, K. (1996). Concept learning using complexity regularization. {\em IEEE Transactions on Information Theory, 42}(1), 48-54.

\im Mayo, D. G. (1996). {\em Error and the Growth of Experimental Knowledge}. University of Chicago Press.


\im Neyman, J. (1977). Frequentist probability and frequentist statistics. {\em Synthese, 36}(1), 97-131.

\im Neyman, J., \& Pearson, E. S. (1933). On the problem of the most efficient tests of statistical hypotheses. {\em Philosophical Transactions of the Royal Society of London. Series A, Containing Papers of a Mathematical or Physical Character, 231}(694-706), 289-337.

\im Neyman, J., \& Pearson, E. S. (1936). Contributions to the theory of testing statistical hypotheses: Part I. {\em Statistical Research Memoirs, 1}, 1-37.

\im Norton, J. D. (2003). A material theory of induction. {\em Synthese, 191}, 671-690.

\im Norton, J. D. (2021). {\em The Material Theory of Induction}. University of Calgary Press.

\im Otsuka, J. (2023). {\em Thinking about Statistics: The Philosophical Foundations}. Routledge.

\im Pearson, E. S. (1962). Some thoughts on statistical inference. {\em The Annals of Mathematical Statistics, 33}(2), 394-403.

\im Popper, K. R. (1959$a$). The propensity interpretation of probability. {\em The British Journal for the Philosophy of Science, 10}(37), 25-42.

\im Popper, K. R. (1959$b$). {\em The Logic of Scientific Discovery}. Hutchinson.

\im Putnam, H. (1965). Trial and error predicates and the solution to a problem of Mostowski. {\em Journal of Symbolic Logic, 30}(1), 49-57.

\item Rubin, D. B. (1974). Estimating causal effects of treatments in randomized and nonrandomized studies. {\em Journal of Educational Psychology 66}: 688-701.

\im Schulte, O. (1999). Means-ends epistemology. {\em The British Journal for the Philosophy of Science, 50}(1), 1-31.

\im Shao, J. (1997). An asymptotic theory for linear model selection. {\em Statistica Sinica, 7}(2), 221-242.

\im Shalev-Shwartz, S., \& Ben-David, S. (2014). {\em Understanding Machine Learning: From Theory to Algorithms}. Cambridge University Press.

\im Shibata, R. (1981). An optimal selection of regression variables. {\em Biometrika, 68}(1), 45-54.

\im Sober, E. (2000). {\em Philosophy of Biology}, 2nd edition. Westview Press.

\im Sober, E. (2008). {\em Evidence and Evolution: The Logic Behind the Science}. Cambridge University Press.

\im Sterkenburg, T. F. (2025). Statistical learning theory and Occam's razor: the core argument. {\em Minds and Machines, 35}(1), 1-28.

\im Swinburne, R. (1997). {\em Simplicity as Evidence of Truth}. Marquette University Press.

\im Vapnik, V., \& Chervonenkis, A. (1974). {\em Theory of Pattern Recognition}. Nauka.

\im von Mises R. (1957) {\em Probability, Statistics and Truth}, revised English edition. New York: Macmillan.


\im Yang, Y. (2005). Can the strengths of AIC and BIC be shared? {\em Biometrika, 92}(4), 937-950.

\im Zabell, S. L. (2005). {\em Symmetry and Its Discontents: Essays on the History of Inductive Probability.} Cambridge University Press.
\end{description}

\end{document}